\documentstyle[12pt]{article}
\begin{document}

\begin{flushright}
{\bf hep-ph/0004120} \\
{\bf LMU-00-04} \\
April 2000
\end{flushright}

\vspace{0.2cm}

\begin{center}
{\large\bf A Novel Possibility to Determine the $CP$-violating Phase $\gamma$ \\
and the $B^0_s$-$\bar{B}^0_s$ Mixing Parameter $y_s$ 
at the $\Upsilon (5S)$ Resonance}
\end{center}

\vspace{.5cm}
\begin{center}
{\bf Zhi-zhong Xing} \footnote{
E-mail: Xing$@$hep.physik.uni-muenchen.de }\\
{\small\sl Sektion Physik, Universit${\sl\ddot a}$t M${\sl\ddot u}$nchen, 
Theresienstrasse 37A, 80333 M${\sl\ddot u}$nchen, Germany}
\end{center}

\vspace{3.5cm}

\begin{abstract}
We show that a $CP$-violating phase can be model-independently
determined from the time-independent measurement of
coherent $B^0_s\bar{B}^0_s$ decays into 
($D^{(*)\pm}_s K^{(*)\mp}$)($D^{(*)\pm}_s K^{(*)\mp}$)
states at the $\Upsilon (5S)$ resonance. This
phase amounts to $(-\gamma)$ within the standard model, where
$\gamma$ is the well-known angle of the quark mixing
unitarity triangle. It is also possible to determine or constrain 
the $B^0_s$-$\bar{B}^0_s$ mixing parameter $y_s$ from the
same measurement.
\end{abstract}

\newpage

Today the weak decays of $B_d$ and $B_s$ mesons 
are playing crucial roles in the study of flavor
mixing and $CP$ violation beyond the neutral kaon system.
The best place to produce $B^0_d$ and $\bar{B}^0_d$ events 
with high statistics and low backgrounds is the 
$\Upsilon (4S)$ resonance, on which both the symmetric $B$
factory at Cornell and the asymmetric $B$ factories at KEK 
and SLAC are based. Similarly 
a wealth of coherent $B^0_s$ and $\bar{B}^0_s$ mesons
can be achieved at the $\Upsilon (5S)$ resonance. 
Recently some interest has been paid to the possibilities to
investigate $CP$ violation and probe new physics in weak $B_s$ decays 
at the $\Upsilon (5S)$ resonance \cite{Xing98,Falk}, 
although it remains an open question whether the existing $B$
factories running at the $\Upsilon (4S)$ energy threshold will finally
be updated to run at the $\Upsilon (5S)$ energy threshold.

\vspace{0.4cm}

In this paper we point out a novel idea, which works
for the coherent decays of $B^0_s \bar{B}^0_s$ pairs into 
$(D^{\pm}_s K^{\mp})(D^{\pm}_s K^{\mp})$,
$(D^{*\pm}_s K^{\mp})(D^{*\pm}_s K^{\mp})$,
$(D^{\pm}_s K^{*\mp})(D^{\pm}_s K^{*\mp})$ and
$(D^{*\pm}_s K^{*\mp})(D^{*\pm}_s K^{*\mp})$ 
states at the $\Upsilon (5S)$ resonance, 
to extract the $CP$-violating phase 
\begin{equation}
\phi \; \equiv \; \arg \left [ \frac{q}{p} \cdot \frac{V_{ub}V^*_{cs}}
{V^*_{cb}V_{us}} \right ] \; ,
%           (1)
\end{equation}
where $q/p$ describes the weak phase of $B^0_s$-$\bar{B}^0_s$ mixing,
and $V_{ub}$, $V_{cs}$, $V_{cb}$ and $V_{us}$ are four elements of 
the Cabibbo-Kobayashi-Maskawa (CKM) quark mixing matrix. Within 
the standard model $q/p = (V^*_{tb}V_{ts})/(V_{tb}V^*_{ts})$
is an excellent approximation, therefore $\phi$ amounts to
$(-\gamma)$ to a good degree of accuracy, where 
\begin{equation}
\gamma \; \equiv \; \arg \left [ - \frac{V^*_{ub}V_{ud}}{V^*_{cb}V_{cd}}
\right ] \; \;
%           (2)
\end{equation}
denotes one inner angle of the well-known CKM unitarity
triangle \cite{PDG98}. So far numerous methods have
been proposed towards a clean determination of $\gamma$ in
the weak decays of $B$ mesons \cite{BB}. 
The advantages of our present approach are remarkable:
(1) it is completely independent of specific models or 
approximate symmetries in treating the relevant hadronic matrix 
elements of $B$ transitions;
(2) its feasibility does not require any time-dependent 
measurement of $B_s$ decays, which is quite difficult due to the
expected rapid rate of $B^0_s$-$\bar{B}^0_s$ oscillation;
and (3) it remains valid to determine $\phi$ and able to
shed light on $\gamma$, even if there exists a kind of yet unknown
new physics in $B^0_s$-$\bar{B}^0_s$ mixing. 

\vspace{0.4cm}

The transitions $B^0_s \rightarrow D^{(*)-}_s K^{(*)+}$ and
$\bar{B}^0_s\rightarrow D^{(*)-}_s K^{(*)+}$ 
occur only through the tree-level 
quark diagrams with the CKM factors $(V^*_{cb}V_{us})$ and
$(V_{ub}V^*_{cs})$, respectively (see Fig. 1 for illustration). 
Their $CP$-conjugate
processes $\bar{B}^0_s\rightarrow D^{(*)+}_s K^{(*)-}$ and
$B^0_s\rightarrow D^{(*)+}_s K^{(*)-}$ have the corresponding
CKM factors $(V_{cb}V^*_{us})$ and $(V^*_{ub}V_{cs})$.
Hence each of the four decay amplitudes involves only a
single weak phase and a single strong phase \cite{ADK}. 
For the study of $CP$ violation it is convenient to define two
rephasing-invariant measurables:  
\begin{eqnarray}
\lambda_{f} & \equiv &
\frac{q}{p} \cdot \frac{\langle f|\bar{B}^0_s\rangle}
{\langle f|B^0_s\rangle} \;\; ,
\nonumber \\
\lambda_{\bar f} & \equiv &
\frac{q}{p} \cdot \frac{\langle \bar{f}|\bar{B}^0_s\rangle}
{\langle \bar{f}|B^0_s\rangle} \;\; ,
%           (3)
\end{eqnarray}
where $f$ and $\bar{f}$ are charge-conjugate states: 
\begin{eqnarray}
f & = & D^-_s K^+ \; , ~ D^{*-}_s K^+ \; , ~ D^-_s K^{*+} \; , ~
D^{*-}_s K^{*+} \; ; \nonumber \\
~\bar{f} & = & D^+_s K^- \; , ~ D^{*+}_s K^- \; , ~ D^+_s K^{*-} \; , ~
D^{*+}_s K^{*-} \; .
%           (4)
\end{eqnarray}
Note that $|q/p| = 1$ holds up to an accuracy of 
${\cal O}(10^{-4})$ in the standard model \cite{Lusignoli}, 
and it is expected to remain valid up to an accuracy of 
${\cal O}(10^{-2})$ even if there exists a kind of yet unknown 
new physics with large $CP$ violation in $B^0_s$-$\bar{B}^0_s$ 
mixing \cite{Xing98,NP}.
We therefore take $|q/p| \approx 1$ as a good approximation in
the subsequent discussions. Then $\lambda_{f}$ and $\lambda_{\bar f}$   
can be explicitly parametrized as \cite{ADK}
\begin{eqnarray}
\lambda_{f} & = & 
\rho ~ e^{i (\phi + \delta)} \; ,
\nonumber \\
\lambda_{\bar f} & = & 
\frac{1}{\rho} ~ e^{i (\phi - \delta)} \; ,
%           (5)
\end{eqnarray}
where $\phi$ is the overall weak phase defined already in Eq. (1), 
$\delta$ is the relevant strong phase difference, and 
\begin{equation}
\rho \; \equiv \; 
\left | \frac{\langle f | \bar{B}^0_s \rangle}
{\langle f | B^0_s \rangle} \right |
\; =\; \left | \frac{\langle \bar{f} | B^0_s \rangle}
{\langle f | B^0_s \rangle} \right | 
%           (6)
\end{equation}
measures the ratio of two decay amplitudes. Of course both $\rho$
and $\delta$ depend upon the specific final state $f$.
%%%%%%%%%%%%%%%%%%%%%%%%%%%%%%%%% Fig. 1 %%%%%%%%%%%%%%%
\begin{figure}
\begin{picture}(400,250)(-20,0)
%---------------------(1)
\put(70,215){\line(1,0){90}}
\put(62,213){$\bar{b}$}
\put(62,186){$s$}
\put(42,200){$B^0_s$}
\put(163,212){$\bar{c}$}
\put(163,186){$s$}
\put(163,244){$u$}
\put(163,228){$\bar{s}$}
\put(180,236){$K^{(*)+}$}
\put(180,199){$D^{(*)-}_s$}
\put(70,190){\line(1,0){90}}
\put(160,240){\oval(70,15)[l]}
\put(145,247.5){\vector(1,0){2}}
\put(145,232.5){\vector(-1,0){2}}
\put(85,215){\vector(-1,0){2}}
\put(145,215){\vector(-1,0){2}}
\put(85,190){\vector(1,0){2}}
\put(145,190){\vector(1,0){2}}
\multiput(110,215)(3,5){5}{\line(0,1){5}}
\multiput(107,215)(3,5){6}{\line(1,0){3}}
%------------------------(2)
\put(280,240){\line(1,0){90}}
\put(280,190){\line(1,0){90}}
\put(272,236){$\bar{b}$}
\put(272,186){$s$}
\put(250,210.5){$B^0_s$}
\put(373,238){$\bar{c}$}
\put(373,187){$u$}
\put(373,225){$s$}
\put(373,200){$\bar{s}$}
\put(386,191){$K^{(*)+}$}
\put(386,230){$D^{(*)-}_s$}
\put(370,215){\oval(70,25)[l]}
\put(295,240){\vector(-1,0){2}}
\put(295,190){\vector(1,0){2}}
\put(355,240){\vector(-1,0){2}}
\put(355,190){\vector(1,0){2}}
\put(355,227.5){\vector(1,0){2}}
\put(355,202.5){\vector(-1,0){2}}
\multiput(310,233.6)(0,-6.1){8}{$>$}
\end{picture}

\begin{picture}(400,250)(-20,-160)
%---------------------(1)
\put(70,215){\line(1,0){90}}
\put(62,213){$b$}
\put(62,186){$\bar{s}$}
\put(42,200){$\bar{B}^0_s$}
\put(163,212){$u$}
\put(163,186){$\bar{s}$}
\put(163,244){$\bar{c}$}
\put(163,228){$s$}
\put(180,236){$D^{(*)-}_s$}
\put(180,199){$K^{(*)+}$}
\put(70,190){\line(1,0){90}}
\put(160,240){\oval(70,15)[l]}
\put(145,247.5){\vector(-1,0){2}}
\put(145,232.5){\vector(1,0){2}}
\put(85,215){\vector(1,0){2}}
\put(145,215){\vector(1,0){2}}
\put(85,190){\vector(-1,0){2}}
\put(145,190){\vector(-1,0){2}}
\multiput(110,215)(3,5){5}{\line(0,1){5}}
\multiput(107,215)(3,5){6}{\line(1,0){3}}
%------------------------(2)
\put(280,240){\line(1,0){90}}
\put(280,190){\line(1,0){90}}
\put(272,236){$b$}
\put(272,186){$\bar{s}$}
\put(250,210.5){$\bar{B}^0_s$}
\put(373,238){$u$}
\put(373,187){$\bar{c}$}
\put(373,225){$\bar{s}$}
\put(373,200){$s$}
\put(386,191){$D^{(*)-}_s$}
\put(386,230){$K^{(*)+}$}
\put(370,215){\oval(70,25)[l]}
\put(295,240){\vector(1,0){2}}
\put(295,190){\vector(-1,0){2}}
\put(355,240){\vector(1,0){2}}
\put(355,190){\vector(-1,0){2}}
\put(355,227.5){\vector(-1,0){2}}
\put(355,202.5){\vector(1,0){2}}
\multiput(310,233.6)(0,-6.1){8}{$>$}
\end{picture}
\vspace{-12cm}
\caption{Quark diagrams for $B^0_s$ and $\bar{B}^0_s$ decays
into $D^-_s K^+$, $D^{*-}_s K^+$, $D^-_sK^{*+}$ or $D^{*-}_sK^{*+}$.}
\end{figure}
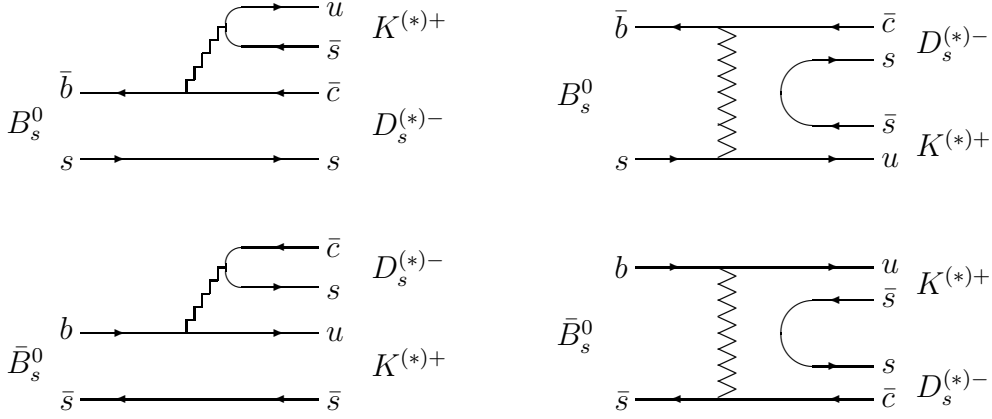
%%%%%%%%%%%%%%%%%%%%%%%%%%%%%%%%%%%%%%%%%%%%%%%%%%%%%%%%%

\vspace{0.4cm}

The coherent $B^0_s$ and $\bar{B}^0_s$ pairs with odd or even 
charge-conjugation ($C$) parity can be produced at the $\Upsilon (5S)$ 
resonance \cite{5S}. The joint decay rates of $B^0_s\bar{B}^0_s$
mesons into $f_1$ and $f_2$ states, in both $C=-1$ and $C=+1$ cases, 
have been derived in Refs. \cite{Xing97,Xing96}. Here we are only interested in
the time-independent measurements.
The generic formula for the time-integrated rate of a joint 
$B^0_s\bar{B}^0_s$ decay mode reads \cite{Xing97}
\begin{eqnarray}
{\cal R}(f_1, f_2)_C & \propto &
|\langle f_1 | B^0_s\rangle|^2 |\langle f_2 | B^0_s \rangle|^2
\left [ \frac{1+Cy^2_s}{(1-y^2_s)^2} \left ( |\xi_C|^2 + |\zeta_C|^2
\right ) - \frac{2 (1+C) y_s}{(1-y^2_s)^2} {\rm Re} \left (\xi^*_C \zeta_C
\right ) \right . \nonumber \\
& & \left . - \frac{1 -C x^2_s}{(1 + x^2_s)^2} \left (|\xi_C|^2 -
|\zeta_C|^2 \right ) +
\frac{2(1+C)x_s}{(1+x^2_s)^2} {\rm Im} \left (\xi^*_C \zeta_C \right )
\right ] \; ,
%           (7)
\end{eqnarray}
where $x_s \equiv \Delta M/\Gamma$ and $y_s \equiv \Delta \Gamma /(2\Gamma)$ 
are two dimensionless parameters of $B^0_s$-$\bar{B}^0_s$ mixing, and
\begin{eqnarray}
\xi_C & = & \frac{p}{q} \left ( 1 + C  \lambda_{f_1} \lambda_{f_2} \right ) 
\; , \nonumber \\
\zeta_C & = & \frac{p}{q} \left (\lambda_{f_2} + C \lambda_{f_1} \right ) \; .
%           (8)
\end{eqnarray} 
The definition for $\lambda_{f_1}$ and $\lambda_{f_2}$ is similar to 
that for $\lambda_{f}$ in Eq. (3).
The present experimental bound for $x_s$ is $x_s \geq 14$ at
the $95\%$ confidence level \cite{PDG98}. In addition, 
a detailed calculation based on the standard model yields 
$y_s \sim {\cal O}(10^{-2})$ up to 0.1 \cite{Beneke}.
This value will always be reduced, if
$B^0_s$-$\bar{B}^0_s$ mixing receives $CP$-violating contributions
from new physics \cite{Xing98,Grossman}. 
Therefore the formula in Eq. (7)  
can be simplified by neglecting the ${\cal O}(y^2_s)$ and
${\cal O}(x^{-2}_s)$ terms. Up to the corrections of ${\cal O}(10^{-2})$, we
arrive for $C= - 1$ at
\begin{equation}
{\cal R}(f_1, f_2)_- \; \propto \;
|\langle f_1 | B^0_s\rangle|^2 |\langle f_2 | B^0_s \rangle|^2
\left [ 1 + |\lambda_{f_1}|^2 + |\lambda_{f_2}|^2 +
|\lambda_{f_1}|^2 |\lambda_{f_2}|^2 - 4 {\rm Re}\lambda_{f_1}
{\rm Re}\lambda_{f_2} \right ] \; ;
%           (9)
\end{equation}
and for $C=+1$ at
\begin{eqnarray}
{\cal R}(f_1, f_2)_+ & \propto &
|\langle f_1 | B^0_s\rangle|^2 |\langle f_2 | B^0_s \rangle|^2
\left [ 1 + |\lambda_{f_1}|^2 + |\lambda_{f_2}|^2 +
|\lambda_{f_1}|^2 |\lambda_{f_2}|^2 + 4 {\rm Re}\lambda_{f_1}
{\rm Re}\lambda_{f_2} ~~~~~ \right . 
\nonumber \\
&  & \left . - 4y_s \left (1 + |\lambda_{f_1}|^2 \right )
{\rm Re} \lambda_{f_2} -4y_s \left ( 1 + |\lambda_{f_2}|^2 \right )
{\rm Re} \lambda_{f_1} \right ] \; . 
%           (10)
\end{eqnarray}
If $y_s \sim {\cal O}(10^{-3})$ held, the relevant ${\cal O}(y_s)$ terms 
in ${\cal R}(f_1, f_2)_+$ could also be neglected. Here and
hereafter we treat $y_s$ as a free parameter of ${\cal O}(10^{-2})$.
It can be seen later on that the formulas in Eqs. (9) and (10) allow one to
determine the weak phase $\phi$ and the mixing parameter $y_s$, 
model-independently, from the joint decays of $B^0_s\bar{B}^0_s$
pairs into $(f, f)$, $(\bar{f}, \bar{f})$ and $(f, \bar{f})$ states.

\vspace{0.4cm}

Now taking $(f_1, f_2) = (f, f)$, $(\bar{f}, \bar{f})$
and $(f, \bar{f})$ respectively, we obtain two
ratios of the three joint decay rates for the $C=-1$ case:
\begin{eqnarray}
R^{(-)}_{f} & \equiv & \frac{{\cal R}(f, f)_-}
{{\cal R}(f, \bar{f})_-} 
\nonumber \\
& = & \frac{(1 + \rho^2 )^2 - 4 \rho^2 \cos^2 (\phi + \delta)}
{(1 + \rho^2 )^2 - 4 \rho^2 \cos (\phi + \delta) \cos (\phi - \delta) } \; \; ,
\nonumber \\ \nonumber \\ 
R^{(-)}_{\bar f} & \equiv & \frac{{\cal R}(\bar{f}, \bar{f})_-}
{{\cal R}(f, \bar{f})_-} 
\nonumber \\
& = & \frac{(1 + \rho^2 )^2 - 4 \rho^2 \cos^2 (\phi - \delta)}
{(1 + \rho^2 )^2 - 4 \rho^2 \cos (\phi + \delta) \cos (\phi - \delta) } \; \; ;
%           (11)
\end{eqnarray}
and another two ratios for the $C=+1$ case:
\begin{eqnarray}
R^{(+)}_{f} & \equiv & \frac{{\cal R}(f, f)_+}
{{\cal R}(f, \bar{f})_+} \; \nonumber \\
& = & \frac{(1 + \rho^2)^2 + 4 \rho^2 \cos^2 (\phi + \delta)
-8y_s\rho (1+\rho^2) \cos (\phi + \delta)}
{(1 + \rho^2)^2 + 4 \rho^2 \cos (\phi + \delta)
\cos (\phi - \delta) - 8y_s\rho (1+\rho^2) \cos \phi \cos\delta} \; \; ,
\nonumber \\ \nonumber \\ 
R^{(+)}_{\bar f} & \equiv & \frac{{\cal R}(\bar{f}, \bar{f})_+}
{{\cal R}(f, \bar{f})_+} \; \nonumber \\
& = & \frac{(1 + \rho^2)^2 + 4 \rho^2 \cos^2 (\phi - \delta)
- 8y_s\rho (1+\rho^2) \cos (\phi - \delta)}
{(1 + \rho^2)^2 + 4 \rho^2 \cos (\phi + \delta) \cos (\phi - \delta) 
- 8y_s\rho (1+\rho^2) \cos\phi \cos\delta} \; \; ;
%           (12)
\end{eqnarray}
The four observables $R^{(\pm)}_{f}$ and $R^{(\pm)}_{\bar f}$ 
depend upon four unknown quantities 
$\rho$, $\phi$, $\delta$ and $y_s$,
therefore the latter can be determined from the former with some
discrete ambiguities (only for $\phi$ and $\delta$). 
Such ambiguities can be resolved if the coherent
$B^0_s\bar{B}^0_s$ decays into four different final states listed
in Eq. (4), which involve the universal $\phi$ and $y_s$ but the 
different $\rho$ and $\delta$, are taken into account.

\vspace{0.4cm}

Note that $\delta = 0$ or $\pi$ would leads to
\begin{equation}
R^{(-)}_{f} \; =\; R^{(-)}_{\bar f} \; =\;
R^{(+)}_{f} \; =\; R^{(+)}_{\bar f} \; =\; 1 \; .
%           (13)
\end{equation}
In this case it is impossible to extract the weak phase $\phi$
from these observables. However,
the four strong phases $\delta (D_sK)$, $\delta (D^*_sK)$, 
$\delta (D_sK^*)$ and $\delta (D^*_sK^*)$ are in general expected
to take values different from one another. Thus one can always 
pin down the magnitude of $\phi$ from the measurements of
$R^{(\pm)}_f$ and $R^{(\pm)}_{\bar f}$ for different final states 
$f$. A special value of $\phi$ ($=0$ or $\pi$) could also result in 
the relationship in Eq. (13), but this possibility should not happen within
the standard model, where $\phi \approx \gamma$ with $0 < \gamma < \pi$. 

\vspace{0.4cm}

An estimate of $\rho$ in the naive factorization approximation
yields $\rho \sim {\cal O}(1)$. Therefore the $B^0_s$-$\bar{B}^0_s$
mixing parameter $y_s$ of ${\cal O}(10^{-2})$ may affect the
magnitudes of $R^{(+)}_{f}$ and $R^{(+)}_{\bar f}$ significantly,
in particular if the $\cos (\phi\pm \delta)$ terms are of
${\cal O}(1)$. This provides an interesting opportunity to determine
or constrain $y_s$ model-independently and time-independently.   

\vspace{0.4cm}

As the branching ratios of $B^0_s$ decays into $D^{\mp}_s K^{\pm}$,
$D^{*\mp}_s K^{\pm}$, $D^{\mp}_s K^{*\pm}$ and $D^{*\mp}_s K^{*\pm}$
states are all at the ${\cal O}(10^{-4})$ level \cite{ADK}, to
observe signals of the joint $B^0_s\bar{B}^0_s$ decays into
these final states needs about $10^8$ $B^0_s\bar{B}^0_s$ events at
the $\Upsilon (5S)$ resonance. The similar number of 
$B^0_s\bar{B}^0_s$ pairs is required to apply the idea proposed 
in Ref. \cite{Falk} for a determination of $\gamma$ using the
partial rates for $CP$-tagged $B_s$ decays into 
$D^{(*)\mp}_s K^{(*)\pm}$ states at the $\Upsilon (5S)$ resonance.

\vspace{0,4cm}

Finally let us give a brief comparison between our new approach 
to determining the $CP$-violating phase $\gamma$ and those 
presented in Ref. \cite{Falk} and Ref. \cite{ADK}, which all make
use of the $B_s$ decays into $D^{(*)\mp}_s K^{(*)\pm}$ states.
The method of Ref. \cite{ADK} requires the time-dependent measurements,
therefore its feasibility relies crucially upon the knowledge of 
$B^0_s$-$\bar{B}^0_s$ mixing (i.e., the known values of $x_s$ and $y_s$). 
To apply the time-independent method of Ref. \cite{Falk}, one
needs the quantitative information on the relevant decay amplitudes 
of {\it pure} $B^0_s$ and $\bar{B}^0_s$ mesons, which can be experimentally
achieved only after the values of $x_s$ and $y_s$ have been fixed. 
For our approach, the precise knowledge of $B^0_s$-$\bar{B}^0_s$
mixing is not a prerequisite. Indeed it is even possible to measure
the magnitude of $y_s$ using our method at the $\Upsilon (5S)$ resonance.
Within the standard model the three approaches can be
complementary to one another for the determination of $\gamma$.

\vspace{0.4cm}

In summary, we have proposed a new method to extract the weak
phase $\gamma$ and to determine the $B^0_s$-$\bar{B}^0_s$ mixing
parameter $y_s$ from the coherent $B^0_s\bar{B}^0_s$ decays into
$(D^{(*)\mp}_s K^{(*)\pm})(D^{(*)\mp}_s K^{(*)\pm})$ states at
the $\Upsilon (5S)$ resonance. It will become realistic and 
useful, perhaps in the second (or final) round of $B$-factory experiments,
once the present $e^+e^-$ colliders operating at the 
$\Upsilon (4S)$ resonance are updated to run at 
the $\Upsilon (5S)$ energy threshold. 

\vspace{0.4cm}

The author is grateful to X. Calmet, A.F. Falk, C.D. L$\rm\ddot{u}$,
and A.A. Petrov for useful communications.

%\newpage

\end{document}